\documentstyle[epsfig,graphics,graphicx,preprint,aps]{revtex}
\begin{document}
\draft
\title{Wyman's solution, self-similarity and 
critical behaviour.} 
\author{G. Oliveira-Neto\thanks{Email:
gilneto@fisica.ufjf.br} and F. I. 
Takakura\thanks{Email:takakura@fisica.ufjf.br}} 
\address{Departamento de F\'{\i}sica,
Instituto de Ciencias Exatas,
Universidade Federal de Juiz de Fora,
CEP 36036-330, Juiz de Fora,
Minas Gerais, Brazil.}
\date{\today}
\maketitle

\begin{abstract}
We show that the Wyman's solution may be obtained from the
four-dimensional Einstein's equations for a spherically 
symmetric, minimally coupled, massless scalar field by using 
the continuous self-similarity of those equations. The Wyman's
solution depends on two parameters, the mass $M$ and the scalar
charge $\Sigma$. If one fixes $M$ to a positive value, say $M_0$,
and let $\Sigma^2$ take values along the real line we show that 
this solution exhibits critical behaviour. For $\Sigma^2 >0$ the
space-times have eternal naked singularities, for $\Sigma^2 =0$
one has a Schwarzschild black hole of mass $M_0$ and finally 
for $-M_0^2 \leq \Sigma^2 < 0$ one has eternal bouncing 
solutions.
\end{abstract}

\section{Introduction.}
\label{sec:introduction}

The four dimensional space-time generated by a minimally coupled, 
spherically symmetric, static, massless scalar field has been
studied by many authors 
\cite{bergman,wyman,roberts,jetzer,clayton}. 
The general solution was found by M. Wyman in \cite{wyman}. 
From a particular case of the general solution, M. D. Roberts 
showed how to construct a time dependent solution \cite{roberts}. 
This solution has an important physical interest because it may 
represent the gravitational collapse of a scalar field. Later, 
P. Brady and independently Y. Oshiro et al. \cite{brady} showed 
that the Roberts' solution could be derived from the 
appropriated, time-dependent, Einstein-scalar equations by using 
a continuous self-similarity. They also showed that this solution 
exhibits a critical behaviour qualitatively identical to the one 
found numerically by M. W. Choptuik \cite{choptuik}, studying the 
same system of equations.

Using a discrete self-similarity of the system, Choptuik 
explicitly showed that, the collapse results in two families of 
solutions depending on the value of a certain parameter $p$. $p$ 
characterizes the scalar charge. In the first family, when 
$p < p_c$, the scalar field collapses up to a certain surface 
and then disperses. In the other family, when $p > p_c$, the 
scalar field collapses to form a black hole. The critical value 
$p_c$ separates the two end states of the collapse. The critical 
solution ($p = p_c$) represents a naked singularity. Therefore, 
since it is given by a single value of the parameter, it has zero 
measure in the parameter space. In fact, the above results 
confirmed early studies of D. Christodoulou who pioneered 
analytical studies of that model \cite{chris}.

The Wyman's solution is not usually thought to be of great 
importance for the issue of gravitational collapse because it
is static and the naked singularities derived from it are 
unstable against spherically symmetric linear pertubations of 
the system \cite{jetzer,clayton}. On the other hand, as we saw 
above, from a particular case of this solution one may derive 
the Roberts' one which is of great importance for the issue of 
gravitational collapse. Also, it was shown that there are 
nakedly singular solutions to the static, massive scalar field 
equations which are stable against spherically symmetric linear 
pertubations \cite{clayton}. Therefore, we think it is of great
importance to gather as much information as we can about the
Wyman's solution for they may be helpful for a better understand
of the scalar field collapse.

In the present work, we would like to show that the Wyman's 
solution may be obtained from the four-dimensional Einstein's 
equations for a spherically symmetric, minimally coupled,
massless scalar field by using the continuous self-similarity 
of those equations. The way by which we shall re-write the 
Einstein-scalar equations in terms of a single variable is
different from the one used in the derivation of the Roberts' 
solution in \cite{brady}. We use a coordinate system with two 
null coordinates $(u,v)$ and the usual angular coordinates. 
Due to the spherical symmetry all the functions appearing in
the field equations must depend only on $u$ and $v$. Our 
system is characterized by three functions: $\sigma$ and $r$, 
present in the metric and the scalar field $\Phi$. Using the
continuous self-similarity of the equations we re-write all
functions in terms of a single variable $z\equiv v/u$, in the
following way: $\exp(2\sigma) \to - f,_u f,_v$ and $f(u,v) \to 
f(z)$, where $f(u,v)$ is an auxiliary function; $r(u,v) \to 
r(z)$; and $\Phi(u,v) \to \Phi(z)$. Therefore, the difference
between the way the authors in \cite{brady} re-write the 
Einstein-scalar equations in terms of a single variable and 
ours, is due to the $\sigma$ and $r$ transformations. We
believe that the above described way, may be sistematically 
used in other self-similar system of equations. After that,
we find the differential equations for the variables $r$,
$\sigma$, $\Phi$ and show that the Wyman's solution, in the
appropriate coordinates, satisfies them.

The Wyman's solution depends on two parameters, the mass $M$ 
and the scalar charge $\Sigma$. As we have seen above, 
Choptuik found a critical behaviour in the solutions to the 
Einstein-scalar equations in terms of the parameter describing
the scalar charge. Therefore, one may try to find critical
behaviour in Wyman's solution by fixing $M$ and letting 
$\Sigma$ varies. We would like to show that, if one fixes $M$ 
to a positive value, say $M_0$, and let $\Sigma^2$ take values 
along the real line, this solution exhibits critical behaviour. 
For $\Sigma^2 >0$ the space-times have eternal naked 
singularities, for $\Sigma^2=0$ one has a Schwarzschild black 
hole of mass $M_0$ and finally for $-M_0^2 \leq \Sigma^2 < 0$ 
one has eternal bouncing solutions. Here, we can see that the 
critical solution is a Schwarzschild black hole which 
interpolates between an infinity number of eternal naked 
singularities and bouncing space-times. Although, this 
behaviour is very different from the one discovered by 
Choptuik, one may have other types of critical solutions in 
the Einstein-scalar system, as demontrated by P. Brady in 
\cite{brady1}.

In the next section, Section \ref{section:self-similarity},
we show that the Wyman's solution may be obtained from the
four-dimensional Einstein's equations for a spherically 
symmetric, minimally coupled, massless scalar field by using 
the continuous self-similarity of those equations. 

In Section \ref{section:critical}, we show that the Wyman's 
solution exhibits critical behaviour. In particular, we
numerically solve, with the appropriate boundary conditions, 
the equation for the radial function ($r$) in terms of the 
radial coordinate ($R$) and show how this function behaves 
for each type of solution: eternal naked singularities,
eternal bouncing solutions and the Schwarzschild black hole. 

Finally, in Section \ref{section:conclusion}, we summarize 
the main points and results of the paper.

\section{Wyman's solution and continuous self-similarity.}
\label{section:self-similarity}

We shall start by writing down the ansatz for the space-time 
metric. As we have mentioned before, we would like to determine 
the space-time generated by a spherically symmetric, minimally 
coupled, self-similar, collapse of a massless scalar field in 
four-dimensions. Therefore, we shall write our metric ansatz as,

\begin{equation}
\label{1}
ds^2\, =\, -\, 2 e^{2\sigma(u,v)} du dv\, +\, r^2(u,v) 
d\Omega^2\, ,
\end{equation}
where $\sigma(u,v)$ and $r(u,v)$ are two arbitrary functions 
to be determined by the field equations, $(u,v)$ is a pair of 
null coordinates varying in the range $(-\infty,\infty)$, and 
$d\Omega^2$ is the line element of the unit sphere.

The scalar field $\Phi$ will be a function only of the two 
null coordinates and the expression for its stress-energy 
tensor $T_{\alpha\beta}$ is given by \cite{wheeler},

\begin{equation}
\label{2}
T_{\alpha\beta}\, =\, \Phi,_\alpha \Phi,_\beta\, -\,
{1\over 2} g_{\alpha\beta} \Phi,_\lambda \Phi^{,_\lambda}\, .
\end{equation}
where $,$ denotes partial differentiation.

Now, combining Eqs. (\ref{1}) and (\ref{2}) we may obtain the
Einstein's equations which in the units of Ref. \cite{wheeler}
and after re-scaling the scalar field, so that it absorbs the 
appropriate numerical factor, take the following form,

\begin{equation}
\label{3}
2 ( r r,_{uv}\, +\, r,_u r,_v )\, +\, e^{2\sigma}\, =\, 0\, ,
\end {equation}
\begin{equation}
\label{4}
2 r r,_{vv}\,-\, 4 r r,_v \sigma,_v\, =\, -\, r^2 
(\Phi,_v)^2\, ,
\end{equation}
\begin{equation}
\label{5}
2 r r,_{uu}\,-\, 4 r r,_u \sigma,_u\, =\, -\, r^2 
(\Phi,_u)^2\, ,
\end{equation}
\begin{equation}
\label{6}
2 ( r^2 \sigma,_{uv}\, +\, r r,_{uv} )\, =\, -\, r^2 
( \Phi,_u \Phi,_v )\, ,
\end{equation}
The equation of motion for the scalar field, in these 
coordinates, is

\begin{equation}
\label{7}
r \Phi,_{uv}\, +\, r,_u \Phi,_{v}\, +\, r,_v \Phi,_{u}\,
=\, 0\, .
\end{equation}

The above system of non-linear, second-order, coupled, 
partial differential equations (\ref{3})-(\ref{7}) can be 
solved if we explore the fact that it is continuously 
self-similar. More precisely, the solution assumes the 
existence of an homothetic Killing vector of the form,

\begin{equation}
\label{8}
\xi\, =\, u {\partial \over \partial u}\, +\, v 
{\partial \over \partial v}\, ,
\end{equation}
Following Coley \cite{coley}, equation (\ref{8}) 
characterizes a self-similarity of the first kind. We 
can express the solution in terms of the variable 
$z = v/u$. 

In order to obtain our solution, we shall re-write
the above equations (\ref{3}-\ref{7}) in terms of $z$,
in a slightly different way than previous works 
\cite{brady}. We shall write the unknown functions as: 
$\Phi(u,v)=\Phi(z)$, $\sigma(u,v)=\sigma(z)$ and 
$r(u,v)=r(z)$. The system of equations (\ref{3}-\ref{7}) 
will have the mentioned self-similarity if we re-write 
$e^{2\sigma}$ in (\ref{1}) in the following way,

\begin{equation}
\label{9}
e^{2\sigma}\, =\, -\, f,_u f,_v\, ,
\end{equation}
where $f = f(u,v)$ is an arbitrary function of the
null coordinates $u$ and $v$ and we must also write it
as: $f(u,v)=f(z)$. It is clear that the main difference 
between our work and previous ones is in the way we 
re-write $\sigma(u,v)$ and $r(u,v)$ as functions of $z$.

In terms of $z$ and taking in account (\ref{9}), the 
system of equations (\ref{3}-\ref{7}) becomes,

\begin{equation}
\label{10}
2 r ( \dot{r}\, +\, z\ddot{r} )\, +\, 2 z ( \dot{r} )^2\, 
-\, z F^2\, =\, 0\, ,
\end{equation}
\begin{equation}
\label{11}
2 z \ddot{r}\, -\, {2 \dot{r}\over F} ( F\, +\, 2 z 
\dot{F} )\, =\, -\ z r ( \dot{\Phi} )^2\, ,
\end{equation}
\begin{equation}
\label{12}
2 z \ddot{r}\, +\, 2 \dot{r}\, +\, {2 r\over F^2} 
[ F \dot{F}\, +\, z F \ddot{F}\, -\, z ( \dot{F} )^2 ]\, 
=\, -\ z r ( \dot{\Phi} )^2\, ,
\end{equation}
\begin{equation}
\label{13}
(z r^2 \dot{\Phi} )^\cdot\, =\, 0\, ,
\end{equation}
where $\cdot$ means differentiation with respect to $z$,
$\dot{f}(z) \equiv F(z)$ and equations (\ref{4}) and 
(\ref{5}) reduce to equation (\ref{11}).

The above system, equations (\ref{10}-\ref{13}), can
be greatly simplified if we re-write it in terms of
the new variable $2R \equiv \ln z$. The resulting
equations are,

\begin{equation}
\label{14}
(r^2)''\, =\, 4 e^{4 R} F^2\, ,
\end{equation}
\begin{equation}
\label{15}
2 r''\, -\, 4 r' \left( 2\, +\, 
{F'\over F} \right)\, =\, -\, r (\Phi')^2\, ,
\end{equation}
\begin{equation}
\label{16}
2 r''\, +\, 2 r \left( {F'\over F} 
\right)'\, =\, -\, r (\Phi')^2\, ,
\end{equation}
\begin{equation}
\label{17}
( r^2 \Phi' )'\, =\, 0\, ,
\end{equation}
where the $'$ means differentiation with respect
to $R$.

We solve the system of equations (\ref{14}-\ref{17}) 
by initially writing down a second order differential 
equation for $r$ in the following way. First of all,
we subtract equation (\ref{15}) from equation 
(\ref{16}) and manipulate it in order to find,

\begin{equation}
\label{18}
2\, +\, {F'\over F}\, =\, {C\over r^2}\, ,
\end{equation}
where $C$ is a real integration constant. 
Then, we differentiate equation (\ref{14}) once with 
respect to $R$ and introduce, in the differentiated 
equation, the value of $e^{4 R} F^2$ from equation 
(\ref{14}) and the information coming from equation 
(\ref{18}). Finally, we integrate the resulting 
equation to find,

\begin{equation}
\label{19}
r^3 r''\, -\, 2 C r r'\, =\, B\, ,
\end{equation}
where $B$ is a real integration constant. Later, when 
we show that our solution is indeed the Wyman's one, 
we shall identify the physical meaning of $C$ and $B$. 

The equation for $\Phi$ can be obtained if we
eliminate $F$ and its derivatives from either one of
the equations (\ref{15}) or (\ref{16}), with the aid
of equation (\ref{18}). They give the same result,
which is,

\begin{equation}
\label{20}
\Phi'\, =\, {\sqrt{-2B}\over r^2}\, .
\end{equation}
It is easy to see that equation (\ref{17}) also leads
to the same result above. Another way to write $\Phi$,
which will be very important to our work, is derived 
from equation (\ref{20}) with the aid of equation
(\ref{18}). After some algebra and an integration,
we get,

\begin{equation}
\label{21}
\Phi\, =\, {\sqrt{-2B}\over 2 C} \ln (2e^{4R} F^2)\, ,
\end{equation}
where we set the integration constant to $(1/2)\ln 2$.

Therefore, a solution to the system of equations 
(\ref{14}-\ref{17}) is computed by initially solving
equation (\ref{19}). Then, for the function $r(R)$ 
obtained, one uses equation (\ref{14}) to find 
$e^{2\sigma}$. Finally, for the function $r(R)$ obtained, 
one uses equation (\ref{20}) to find the scalar field 
$\Phi(R)$ or equivalently for the value of $e^{4R} F^2$ 
obtained, one uses equation (\ref{21}) to find $\Phi(R)$.

Now, we would like to demonstrate that the Wyman's 
solution is a solution to the system of equations 
(\ref{14}-\ref{17}). As the first step in our 
demonstration, we apply the following coordinate 
transformations,

\begin{equation}
\label{22}
U\, =\, \ln u\, ,\qquad  V\, =\, \ln v\, .
\end{equation}
From the definition of $2R$ as $\ln z$, we immediately
obtain from equation (\ref{22}) that, $2R = V - U$. In
terms of these new coordinates, the line element 
equation (\ref{1}) becomes,

\begin{equation}
\label{23}
ds^2\, =\, -\, 2 e^{4R} F(R)^2 dV dU\, +\, r(R)^2
d\Omega^2\, .
\end{equation}

We may now compare it with the line element of the
Wyman's solution written in the doubel null 
coordinates (U,V) \cite{roberts},

\begin{equation}
\label{24}
ds^2\, =\, -\, \left( 1 - {2\eta \over {\mathcal{R}} }
\right)^{M/\eta} dV dU\, +\,  \left( 1 - {2\eta \over 
{\mathcal{R}} }\right)^{1 - M/\eta }{{\mathcal{R}} }^2 
d\Omega^2\, ,
\end{equation}
where $M$ is the mass parameter, $\eta^2 = M^2 + 
\Sigma^2$ and $\Sigma$ is the scalar charge. The
function ${\mathcal{R}}(U,V)$ may be written in terms
of $R$ in the following way,

\begin{equation}
\label{25}
R\, -\, R_0\, =\, {1\over 1 + M/\eta }\left[ \left( 
1 - {2\eta \over {\mathcal{R}}}\right)^{-M/\eta} 
{\mathcal{R}} \left( 1 - {{\mathcal{R}} \over 
2\eta }\right)^{M/\eta} 
\mbox{\scriptsize 2}F\mbox{\scriptsize 1}\left( 1 + 
{M\over \eta }, {M\over \eta }; 2 + {M\over \eta }; 
{{\mathcal{R}} \over 2\eta } \right) \right]
\end{equation}
where $\mbox{\scriptsize 2}F\mbox{\scriptsize 1}
(a,b;c;z)$ is the Gauss hypergeometric function and 
$R_0$ is an integration constant. This is a 
transcendental equation which cannot be inverted,
in general, to give us $\mathcal{R}$ as an explicity
function of $R$. Nevertheless, we shall consider
$\mathcal{R}$ as an implicity function of $R$ and
use equation (\ref{25}) to prove our thesis.

Now, since $r$ and $\mathcal{R}$ are functions of $R$,
we may directly compare the metric components in the
line elements (\ref{23}) and (\ref{24}) to obtain,

\begin{equation}
\label{26}
r(R)^2\, =\, \left( 1 - {2\eta \over {\mathcal{R}}(R)
}\right)^{1 - M/\eta }{{\mathcal{R}}(R) }^2
\end{equation}
and
\begin{equation}
\label{27}
2 e^{4R} F(R)^2\, =\, \left( 1 - {2\eta \over 
{\mathcal{R}}(R) }\right)^{M/\eta}
\end{equation}
Therefore, from the Wyman's solution we have an
expression for $r$, which must satisfies equation
(\ref{19}) and an expression for 
$e^{2\sigma}=e^{4R}F(R)^2$, which must satisfies 
equation (\ref{14}).

Introducing the expression of $r$ in terms of 
$\mathcal{R}$ (\ref{26}) in equation (\ref{19}), we
may prove, with the aid of equation (\ref{25}), that
it satisfies equation (\ref{19}) if, and only if,
$C=M$ and $-B=\Sigma^2$. From now on, we shall 
assume that those are the correct values of the
integration constants $C$ and $B$. Likewise, if we 
introduce the expressions of $r$ equation (\ref{26}) 
and $e^{2\sigma}$ equation (\ref{27}) in equation 
(\ref{14}), we may prove, with the aid of equation 
(\ref{25}), that this equation is satisfied. We
complete our demonstration by using the value of
$e^{4R}F(R)^2$ equation (\ref{27}), in equation
(\ref{21}) to find,

\begin{equation}
\label{28}
\Phi\, =\, {\Sigma\over 2\eta} \ln \left( 1\, -\,
{2\eta\over {\mathcal{R}}}\right)\, .
\end{equation}
Which is exactly the expression for the scalar field
in the Wyman's solution \cite{roberts}.

So, the Wyman's solution may be derived from the
Einstein-scalar equations (\ref{3}-\ref{6}) by
using the continuous self-similarity of those 
equations in an appropriated way. As a matter of 
completeness we mention that, if we introduce a time 
coordinate $T$ defined by, $2T = V + U$, we 
may re-write the line element equation (\ref{1}), as,

\begin{equation}
\label{29}
ds^2\, =\, 2 e^{4R} F(R)^2 (- dT^2\, +\, 
dR^2)\, +\, r^2(R) d\Omega^2\, .
\end{equation}

\section{Wyman's solution and critical behaviour.}
\label{section:critical}

Depending on the value of the parameters $M$ and 
$\Sigma$ the Wyman's solution may represent 
different static, asymptotically flat space-times 
\cite{bergman}, \cite{roberts}. When one sets 
$\Sigma = 0$, the scalar field vanishes from equation
(\ref{28}) and one obtains the Schwarzschild solution 
with a mass $M$. Therefore, it is usual to consider 
$M$ positive. For positive $M$ one may let $\Sigma^2$ 
take values over the real line. For $\Sigma^2 > 0$,
the solution represents space-times with an eternal 
naked time-like singularity located at $r=0$. From 
equation (\ref{28}), the scalar field vanishes 
asymptotically ($r \to \infty$) and diverges at the 
singularity. For $\Sigma^2 < 0$, one normaly considers 
the domain: $-M^2 \leq \Sigma^2 < 0$ because if 
$\Sigma^2 < -M^2$ from equation (\ref{24}) the metric 
gets complex. The case $\Sigma^2 = -M^2$ is well know 
in the literature as the Yilmaz-Rosen space-time 
\cite{rosen}. Observe that this restriction in the
domain of $\Sigma^2$, when this quantity is negative, 
comes from the use of the quantities $(U, V, 
{\mathcal{R}})$ equation (\ref{24}). When one considers 
the metric written in the original quantities in the 
Wyman's papers \cite{wyman} or even in the quantities 
leadind to equation (\ref{23}), one sees that the 
domain of $\Sigma^2 < 0$ may be extended to: $(0, 
-\infty )$. In the present situation, since we are 
using the identifications (\ref{26}) and (\ref{27}), we 
can see that our original quantities $(U,V,r)$ become 
ill defined for $\Sigma^2 < -M^2$. Therefore, we shall 
consider here the domain of $\Sigma^2$ to be $[ -M^2,
\infty )$. In these space-times, $r$ is never zero. 
If one starts with a large value of $r$ it diminishes, 
as we let it varies as a function of $R$, until it 
reaches a minimum value. Then, it starts to increase 
again without limit. We may interpret this case as an 
eternal bouncing solution. An important property of this 
space-time is that the scalar field equation (\ref{28}) 
is imaginary. The imaginary scalar field also known as
ghost Klein-Gordon field \cite{hayward} is an example
of the type of matter called {\it exotic} by some
authors \cite{thorne}. It violates most of the energy 
conditions and is repulsive. This property helps 
explaing the reason why the collapsing scalar field 
bounces back without reaching $r=0$. Recently, the 
exotic matter has been in evidence due to the discovery
that the universe is expanding in an accelerated rate
\cite{norbert}. This implies that the universe must be 
filed with matter which violates at least the strong 
energy condition \cite{norbert}.
This type of matter is also very much used as the
matter responsible for the formation and maintenance
of traversible wormholes \cite{thorne}, \cite{hayward1}.
In particular, the ghost Klein-Gordon field has been
used as one of the first specific {\it exotic} matter
models to explain the formation and maintenance of
traversible wormholes \cite{hayward}. Besides the
above motivations for the use of {\it exotic} matter,
which involve classical fields, one must not forget
about negative energy densities very commom in quantum
field theories \cite{davies}.

Based on the above properties of the Wyman's solution
and taking in account the results of references 
\cite{brady}, \cite{choptuik}, \cite{chris}, on 
critical behaviour in the spherical symmetric, massless, 
scalar field collapse, we conclude that the Wyman's 
solution also shows a critical behaviour. Although  the
critical solution is of a different type from the one
discovered in the above mentioned refences, one may
proceed here in the same way as the authors there, in
order to show the critical behaviour. One fixes the 
mass parameter $M$ and considers the scalar charge 
$\Sigma$ as the only free parameter. In the present 
case, as discussed above, we fix $M$ to a positive 
value, say $M_0$. $\Sigma^2$ may take values in the
domain, $(-M_0^2 , \infty )$. For  $\Sigma^2$ positive
$(0 < \Sigma^2 < \infty)$, we have a set ($N$) of 
solutions representing asymptotically flat, time-like, 
eternal naked singularities. Consider, now, equation 
(\ref{19}), suitably written for the present situation 
as,

\begin{equation}
\label{30}
r^3 r''\, -\, 2 M_0 r r'\, =\, -\, \Sigma^2 \, .
\end{equation}

We may numerically solve it using boundary conditions 
reflecting the asymptotic behaviour of the solution and
the appropriate value of $\Sigma^2$. With this numerical 
solution, we may draw the curve $r \times R$. Following 
the above procedure, we show in Figure 1 the curve $r 
\times R$ for two members of the set $N$. For $\Sigma^2$ 
negative $(-M_0^2 < \Sigma^2 < 0)$, we have a set ($Bo$)
of solutions representing asymptotically flat, eternal 
bouncing space-times. Figure 2 shows the curve $r \times 
R$ for two members of the set $Bo$, obtained from the 
numerical solutions of equation (\ref{30}) with 
appropriate values of $\Sigma^2$. Finally, for $\Sigma^2 
= 0$, we obtain the critical solution which interpolates 
between the infinity number of solutions in the set $N$ 
and the ones in the set $Bo$. For fix $M$ this solution 
is a point in the parameter space of solutions and 
represents the Schwarzschild black hole with mass $M_0$. 
We may solve equation (\ref{30}) for the appropriate 
value of $\Sigma^2$ and draw the curve $r \times R$ for 
the Schwarzschild black hole. This curve is shown in 
Figure 3.

%%%%%%%%%%%%%%%%%%%%%%
\begin{figure}
\begin{tabular}[c]{ccc}%
\mbox{\hspace{-0.2cm}} & (a) & (b) \\
\mbox{\hspace{-0.2cm}} & 
\resizebox{!}{5cm}{\includegraphics{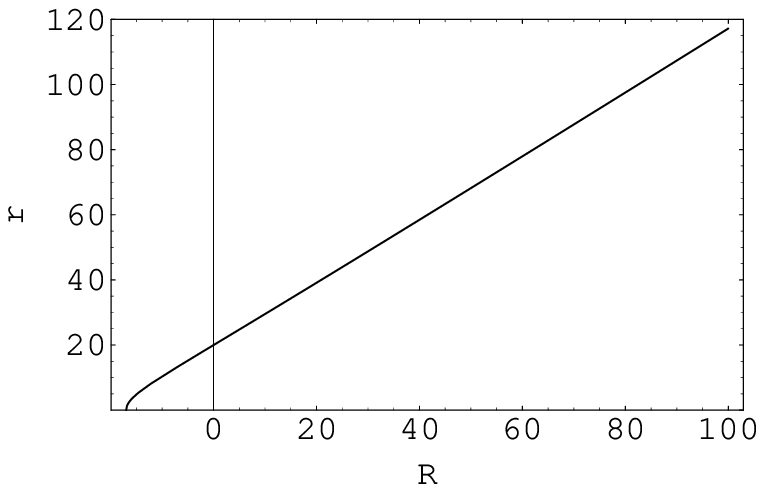}} &
\resizebox{!}{5cm}{\includegraphics{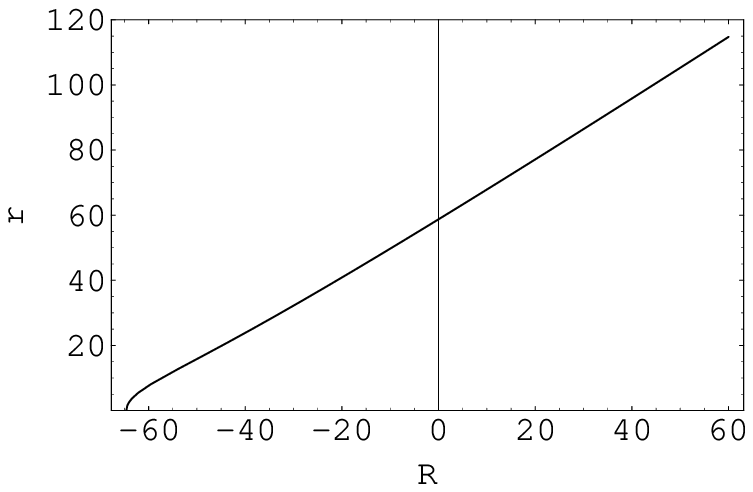}}%
\end{tabular}
\caption{Graphics of $r \times R$ for the eternal naked
singularities with different values of $M_0 \mbox{ and } 
\Sigma^2$. (a) $M_0 = 1$ and $\Sigma^2 = 160$, (b) $M_0 
= 3$ and $\Sigma^2 = 280$. }%
\end{figure}
%%%%%%%%%%%%%%%%%%%%%%
\vskip.5in

%%%%%%%%%%%%%%%%%%%%%%
\begin{figure}
\begin{tabular}[c]{ccc}%
\mbox{\hspace{-0.2cm}} & (a) & (b) \\
\mbox{\hspace{-0.2cm}} & 
\resizebox{!}{5cm}{\includegraphics{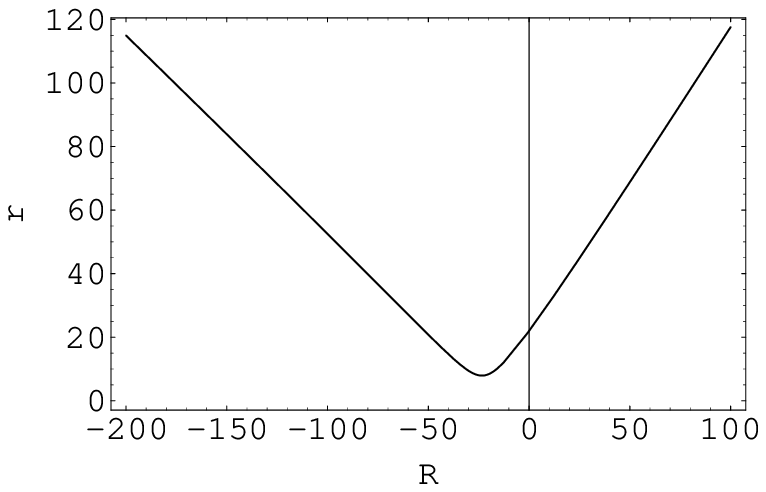}} &
\resizebox{!}{5cm}{\includegraphics{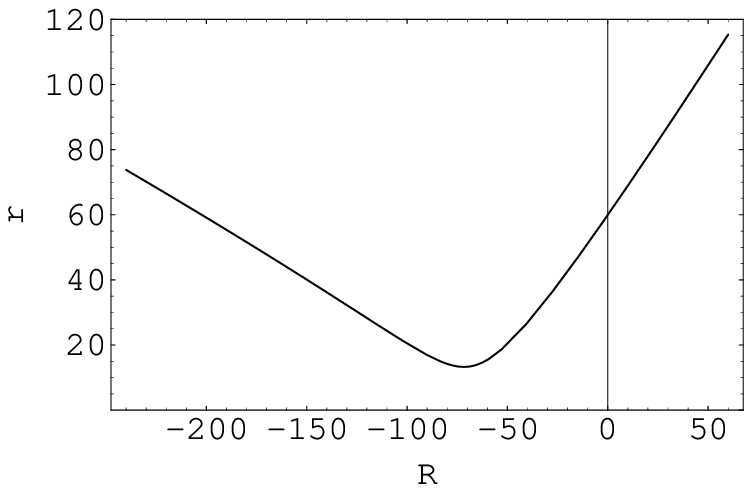}}%
\end{tabular}
\caption{Graphics of $r \times R$ for the eternal bouncing 
space-times with different values of $M_0 \mbox{ and } 
\Sigma^2$. (a) $M_0 = 1$ and $\Sigma^2 = - 160$, (b) $M_0 
= 3$ and $\Sigma^2 = - 280$. }%
\end{figure}
%%%%%%%%%%%%%%%%%%%%%%
\vskip.5in

%%%%%%%%%%%%%%%%%%%%%%
\begin{figure}
\begin{tabular}[c]{ccc}%
\mbox{\hspace{-0.2cm}} & (a) & (b) \\
\mbox{\hspace{-0.2cm}} & 
\resizebox{!}{5cm}{\includegraphics{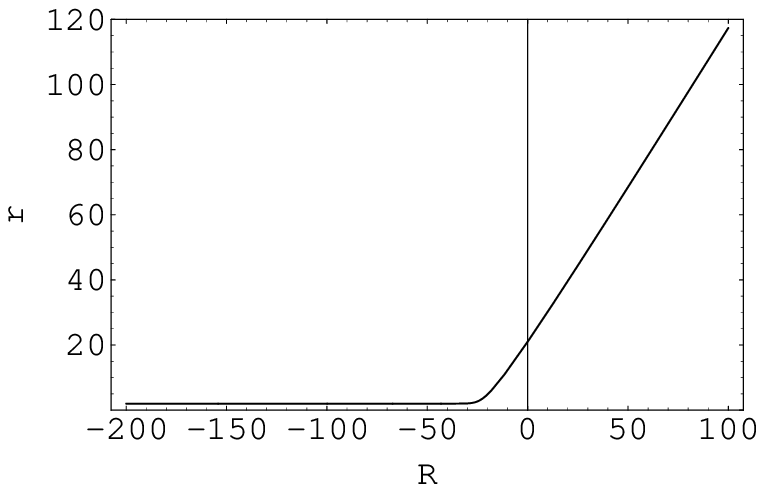}} &
\resizebox{!}{5cm}{\includegraphics{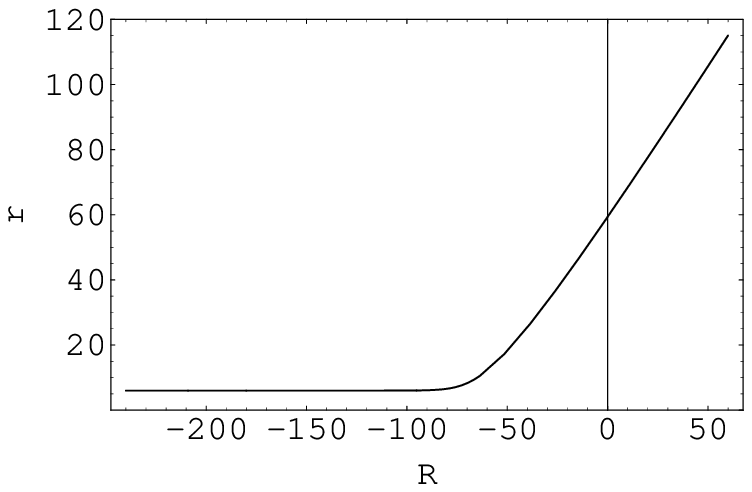}}%
\end{tabular}
\caption{Graphics of $r \times R$ for the Schwarzschild 
space-times with different values of $M_0$. (a) $M_0 = 1$ 
(b) $M_0 = 3$. }%
\end{figure}
%%%%%%%%%%%%%%%%%%%%%%
\vskip.5in

\section{Conclusions.}
\label{section:conclusion}

In the present work, we showed that the Wyman's solution 
may be obtained from the four-dimensional Einstein's 
equations for a spherically symmetric, minimally coupled, 
massless scalar field by using the continuous 
self-similarity of those equations. We also showed that 
the Wyman's solution exhibits critical behaviour. We did 
that by fixing the mass $M$ to a positive value, say 
$M_0$, and letting $\Sigma^2$ takes values along the real 
line. For $\Sigma^2 >0$ the space-times have eternal naked 
singularities, for $\Sigma^2 =0$ one has a Schwarzschild 
black hole of mass $M_0$ and finally for $-M_0^2 \leq 
\Sigma^2 < 0$ one has eternal bouncing solutions. Here, 
the critical solution is a Schwarzschild black hole which 
interpolates between an infinity number of eternal naked 
singularities and bouncing space-times. Although, this 
behaviour is very different from the one discovered by 
Choptuik, one may have other types of critical solutions 
in the Einstein-scalar system, as demonstrated by P. 
Brady in \cite{brady1}.

\acknowledgements

We would like to thank FAPEMIG for the invaluable 
financial support.

\end{document}